\documentclass[prd,amsfonts,nofootinbib,preprintnumbers,showpacs]{revtex4}
\usepackage{amssymb, amsmath, amscd}
\usepackage{graphics}
\usepackage{epsfig}

\newcommand{\Det}{\mathop{\rm Det}\nolimits}
\newcommand{\tr}{\mathop{\rm tr}\nolimits}
\newcommand{\Tr}{\mathop{\rm Tr}\nolimits}
\newcommand{\bra}[1]{\langle#1|}
\newcommand{\ket}[1]{|#1\rangle}

\begin{document}

\title{Gravitational catalysis of chiral
and color symmetry breaking \\ of quark matter in hyperbolic space}

\author{D.~Ebert$^{1,2}$\footnote{debert@physik.hu-berlin.de},
  A.~V.~Tyukov$^{3}$, and
  V.~Ch.~Zhukovsky$^{3}$\footnote{zhukovsk@phys.msu.ru}}

\affiliation{$^{1}$ Joint Institute for Nuclear Research, Dubna,
R-141980, Russia}

\affiliation{$^{2}$ Institut f\"ur Physik, Humboldt-Universit\"at zu
Berlin, 12489 Berlin, Germany}

\affiliation{$^{3}$ Faculty of Physics, Department of Theoretical
Physics, Moscow State University, 119991, Moscow, Russia}

\begin{abstract}
We study the dynamical breaking of chiral and color symmetries of
dense quark matter in
the ultrastatic
hyperbolic
spacetime $R\otimes H^3$
in the framework of an extended Nambu--Jona-Lasinio model. On the
basis of analytical expressions for chiral and color condensates as
functions of curvature and temperature, the phenomenon of
dimensional reduction and gravitational catalysis of symmetry
breaking
in strong gravitational field
is demonstrated
in the regime of weak coupling constants.
In the case of strong couplings it is shown that curvature leads to
small corrections to the flat-space values of condensate and thus
enhances the symmetry breaking effects.
Finally, using numerical calculations phase transitions under the
influence of chemical potential and negative curvature are
considered and the phase portrait
of the system
is constructed.
\end{abstract}

\maketitle

\section{Introduction}

Dynamical  breaking of chiral and color symmetries has been
successfully studied within field theories of the
Nambu--Jona-Lasinio (NJL) type with four-fermion interactions
\cite{NJL}. They are also quite useful in describing the physics of
light mesons (see e.g. \cite{MESONS,generalN,kunihiro} and
references therein) and diquarks \cite{Ebert_Kaschluhn, Vogl}.

The possibility for the existence of the color superconductive (CSC)
phase with a nonzero colored diquark condensate was proposed both in
the region of high baryon densities \cite{Ba,Frau,bl} and moderate
densities \cite{Alford,Berges,Schwarz,Alford_Kebrikov,shovk}. In the
framework of NJL models the CSC phase formation has generally been
considered as a dynamical competition between diquark $\langle
qq\rangle$ and usual quark-antiquark condensation $\langle\bar
qq\rangle$. Special attention has been paid to the catalyzing
influence of external fields on chiral symmetry breaking ($\chi$SB)
in the regime of weak coupling \cite{Klimenko,5,6} (constant
magnetic field), \cite{KMV,ebzhu} (chromomagnetic fields) and on the
condensation of diquarks \cite{toki,Ebert_Zhukovsky} (chromomagnetic
fields). In particular, it was demonstrated that in a strong field
the considered symmetry is dynamically broken for an arbitrary weak
attraction between quarks. The physical explanation for this is that
the effect of dynamical symmetry breaking is accompanied by an
effective lowering of dimensionality in strong fields, where the
number of reduced units of dimensions depends on the concrete type
of the background field. Dynamical symmetry breaking in a magnetic
field in spacetimes of dimension higher than four was also
considered in \cite{7}.

For cosmological and astrophysical applications, it is also
interesting to study the influence of spacetime curvature on
symmetry breaking. One of the ways to account for the effects of
gravity is to use the adiabatic expansion of Green's functions in
the vicinity of a fixed point in powers of small curvature (see, for
example, the review \cite{MUTA}). However, since second-order phase
transitions take place in the infrared region, the considered
processes may become sensitive to the global structure of spacetime,
and then one needs exact expressions for the propagators in the
curved spacetime.

In particular, exact solution can be found
for spacetime with high symmetry.
A variety of examples of $\chi$SB in different symmetric spaces both
at zero and finite temperature has been considered in the literature
(see e.g. \cite{MUTA}). One of the well-known examples of spaces
with constant positive curvature is the Einstein universe of the
form $R\otimes S^3$. $\chi$SB at finite temperature and chemical
potential in the static Einstein universe was recently considered in
\cite{huang}. Further investigations of quark matter in this
gravitational background concerning, in particular, diquark and pion
condensation were performed in \cite{E_Zh_Tyu,E_Zh_Kl_Tyu}. There,
the positive curvature was shown to lead to the restoration of
broken symmetries, thus acting in a similar way as the temperature.
Another symmetric space, where exact solution may be found, is the
ultrastatic hyperbolic spacetime $R\otimes H^3$. It was noted that in spaces
with negative curvature the chiral symmetry is broken even at weak
coupling constant (see e.g. \cite{MUTA}). The detailed analysis of
the heat kernel in hyperbolic space showed that the physical reason
for this phenomenon is the effective dimensional reduction for
fermions in the infrared region \cite{gorbar}. Recently, the
combined influence of gravitational and magnetic fields on $\chi$SB
in the special case of the 2D space of negative curvature, i.e., on
the Lobachevsky plane, was considered in \cite{gorbar_gusynin}. As
discussed by these authors, the study of the effects of surface
curvature may be important for some condensed matter systems
concerning, in particular, the quantum Hall effect in graphene
\cite{graf1,graf2}.

The aim of this paper is to study the effects of dynamical breaking
of chiral and/or color symmetries in dense quark matter under the
influence of negative curvature of
ultrastatic
hyperbolic spacetime $R\otimes H^3$. In the framework of an extended
Nambu--Jona-Lasinio model, including $(\bar{q}q)$- and
$(qq)$-interactions, an
exact in curvature
expression for the thermodynamic potential is derived, which
contains all the necessary information about the condensates. Basing
upon the analytical solutions of gap equations for quark and diquark
condensates we show
that
in strong gravitational field
there arises a gravitational catalysis of dynamical symmetry
breaking
and
chiral and color symmetries may be simultaneously broken even for
weakly interacting quarks.
This situation resembles the influence of magnetic or chromomagnetic
fields on symmetry breaking in flat case.
We also
consider
the role of finite temperature and find
that for any fixed value of curvature there exists a
critical
temperature at which chiral and color symmetries become restored.
Moreover, using numerical calculations,
phase transitions under the influence of a chemical potential are
investigated and the phase portrait of the system at zero
temperature is
constructed.
Finally, it is shown that in the strong coupling regime negative
curvature enhances the values of condensates as compared to the flat
case. Our analysis demonstrates that negative curvature acts on
chiral and color condensates
in a way similar to that of a magnetic field in flat spacetime.

\section{The extended NJL model in curved spacetime}
Let us briefly remind the basic definitions necessary for the
description of fermions in curved spacetime. In 4-dimensional curved
spacetime with signature $(+,-,-,-)$, the line element is written as
\[
ds^2=\eta_{\hat a\hat b}e_\mu^{\hat a}e_\nu^{\hat b}dx^\mu dx^\nu.
\]
The gamma-matrices $\gamma_{\mu}$, metric $g_{\mu\nu}$ and the
vierbein $e^{\mu}_{\hat{a}}$, as well as the definitions of the
spinor covariant derivative $\nabla_{\nu}$ and spin connection
$\omega^{\hat{a}\hat{b}}_{\nu}$ are given by the following
relations~\cite{Parker_Toms,brill}:
\begin{eqnarray}
& & \{\gamma_{\mu}(x),\gamma_{\nu}(x)\}=2g_{\mu\nu}(x), \quad
  \{\gamma_{\hat{a}},\gamma_{\hat{b}}\}=2\eta_{\hat{a}\hat{b}}, \quad
  \eta_{\hat{a}\hat{b}}={\rm diag}(1,-1,-1,-1), \nonumber \\
& & g_{\mu\nu}g^{\nu\rho}=\delta^{\rho}_{\mu}, \quad
  g^{\mu\nu}(x)=e^{\mu}_{\hat{a}}(x)e^{\nu \hat{a}}(x), \quad
  \gamma_{\mu}(x)=e^{\hat{a}}_{\mu}(x)\gamma_{\hat{a}}. \label{2}\\
 & &
  \nabla_{\mu}=\partial_{\mu}+\Gamma_{\mu},\quad\Gamma_{\mu}=\frac12\,\omega^{\hat a \hat b}_{\mu}\sigma_{\hat a \hat
  b},
  \quad
   \sigma_{\hat{a}\hat{b}}=\frac{{1}}{4}[\gamma_{\hat{a}},\gamma_{\hat{b}}],
 \nonumber \\
& & \omega^{\hat{a}\hat{b}}_{\mu}=\frac{1}{2}e^{\hat{a}\lambda}
e^{\hat{b}\rho}[C_{\lambda\rho\mu}-C_{\rho\lambda\mu}-C_{\mu\lambda\rho}],
 \quad C_{\lambda\rho\mu}=
e^{\hat{a}}_{\lambda}\partial_{[\rho}e_{\mu]\hat{a}}. \label{3}
\end{eqnarray}
Here, the index $\hat{a}$ refers to the flat tangent space defined
by the vierbein at the spacetime point $x$, and the
$\gamma^{\hat{a}} (\hat a=0,1,2,3)$ are the usual Dirac
gamma-matrices  of Minkowski spacetime. Moreover $\gamma_5$,  is
defined, as usual (see, e.g., \cite{Parker_Toms,bordag,Camporesi}),
i.e. to be the same as in flat spacetime and thus independent of
spacetime variables.

The extended NJL model which includes the $(\bar q q)$- and $(q
q)$-interactions of colored up- and down-quarks can be used to
describe the formation of the color superconducting phase. For the
color group $SU_c(3)$ its Lagrangian takes the form
\begin{equation}
    \label{lagrangian}
    \mathcal{L} =
    \bar{q} \left[ i \gamma^{\mu}\nabla_{\mu} + \mu \gamma^{0} \right] q
    + \frac{G_{1}}{2N_{c}}
    \left[
        \left( \bar{q} q \right)^{2}
        + \left( \bar{q} i \gamma^{5} \vec{\tau} q \right)^{2}
    \right]
    + \frac{G_{2}}{N_{c}}
    \left[
        i \bar{q}_{c} \varepsilon\epsilon^{b} \gamma^{5} q
    \right]
    \left[
        i \bar{q} \varepsilon\epsilon^{b} \gamma^{5} q_{c}
    \right].
\end{equation}
Here, $N_c=3$ is the number of colors, $G_1$ and $G_2$ are coupling
constants (their particular values will be chosen in what follows),
$\mu$ is the quark chemical potential,
$q_c=\textit{C}\bar{q}^t,\bar{q}_c=q^t\textit{C}$\,\, are
charge-conjugated bispinors ($t$ stands for the transposition
operation). The charge conjugation operation is defined with the
help of the operator $\textit{C}=i\gamma^{\hat2}\gamma^{\hat 0}$,
where the flat-space matrices $\gamma^{\hat2}$ and $\gamma^{\hat 0}$
are used (see, e.g., \cite{Parker_Toms}). The quark field $q\equiv
q_{i\alpha}$ is a doublet of flavors and triplet of colors with
indices $i=1,2;\;\alpha=1,2,3.$ Moreover,
$\vec{\tau}\equiv(\tau^1,\tau^2,\tau^3)$ denote Pauli matrices in
the flavor space;
$(\varepsilon)^{ik}\equiv\varepsilon^{ik},\;(\epsilon^{b})^{\alpha\beta}\equiv\epsilon^{\alpha\beta
  b}$
are the totally antisymmetric tensors in the flavor and color
spaces, respectively.

Next, by applying the usual bosonization procedure, we obtain the
linearized version of the Lagrangian~(\ref{lagrangian}) with
collective boson fields $\sigma$, $\vec\pi$  and $\Delta$,
\begin{equation}
    \label{auxilary}
    \tilde{\mathcal{L}} =
    \bar{q} \left[ i \gamma^{\mu}\nabla_{\mu} + \mu \gamma^{0} \right] q
    - \bar{q}\left(\sigma+i\gamma^5\vec{\tau}\vec{\pi}\right)q
    - \frac3{2G_1}(\sigma^2+\vec{\pi}^2)
    - \frac3{G_{2}} \Delta^{*b} \Delta^b -
    \Delta^{*b}\left[i q^t \textit{C} \varepsilon \epsilon^{b}
\gamma^{5} q\right]
    -\Delta^{b}\left[i \bar{q} \varepsilon \epsilon^{b} \gamma^{5} \textit{C}
\bar{q}^t\right].
\end{equation}

The Lagrangians (\ref{lagrangian}) and (\ref{auxilary}) are
equivalent, as can be seen by using the Euler-Lagrange equations for
the bosonic fields, from which it follows that
\begin{equation}
  \Delta^b=-\frac{G_2}{3}iq^t C\varepsilon \epsilon^b\gamma^5 q,\quad
  \sigma=-\frac{G_1}{3}\bar qq,\quad
  \vec \pi=-\frac{G_1}{3} \bar q i\gamma^5\vec\tau q.
\end{equation}
The fields $\sigma$ and $\vec{\pi}$ are color singlets, and
$\Delta^b$ is a color anti-triplet and flavor singlet.

In what follows, it is convenient to consider the effective action
for boson fields, which is expressed through the integral over quark
fields
\begin{equation}
    \exp\left\{\,iS_{\rm eff}(\sigma,\vec\pi, \Delta^b, \Delta^{*b})\right\}=
        N\int[dq][d\bar{q}]\exp\biggl\{\,i\int d^4x\sqrt{-g}\tilde{\mathcal{L}}\biggr\},
\end{equation}
where
\begin{equation}
    S_{\rm eff}(\sigma,\vec{\pi}, \Delta^b, \Delta^{*b})=-\int d^4x\sqrt{-g}
    \left[
           \frac{3(\sigma^2+\vec{\pi}^2)}{2G_1}+\frac{3\Delta^b\Delta^{*b}}{G_2}
    \right] + S_q,
\end{equation}
with $S_q$ standing for the quark contribution to the effective action.

In the mean field approximation, the fields $\sigma$, $\vec{\pi}$,
$\Delta^b$, $\Delta^{*b}$ can be replaced by their ground state
averages: $\langle\sigma\rangle$, $\langle\vec{\pi}\rangle$,
$\langle\Delta^b\rangle$ and $\langle\Delta^{*b}\rangle$,
respectively. Let us choose the following ground state of our model:
$$\langle\Delta^1\rangle\;=\;\langle\Delta^2\rangle\;=\;\langle\vec{\pi}\rangle\;=\;0.$$
If $\langle\sigma\rangle\;\neq\;0$, the chiral symmetry is broken
dynamically, and if
$\langle\Delta^3\rangle\;\neq\;0$,
the color
symmetry is broken. Evidently, this choice breaks the color symmetry
down to the residual group $SU_c(2)$. (In the following we denote
$\langle\sigma\rangle$, $\langle\Delta^3\rangle \;
 \neq\;0$ by letters $\sigma,\;\Delta$.)

The quark contribution has the following form (for more details
see~\cite{Ebert_Zhukovsky})
\begin{equation}
\begin{array}{lll}
\label{effaction}
    S_q(\sigma,\Delta)&=& -i\ln\Det\left[i\hat{\nabla}-\sigma+\mu\gamma^0\right]
    -\frac{i}{2}\ln\Det\left[4|\Delta|^2 + (-i\hat{\nabla}-\sigma+\mu\gamma^0 )
        (i\hat{\nabla}-\sigma+\mu\gamma^0 ) \right].
\end{array}
\end{equation}
Here, the first determinant is over spinor, flavor and coordinate
spaces, and the second one is over the two-dimensional color space
as well, and $\hat{\nabla}=\gamma^\mu\nabla_\mu$.

Note that the effective potential of the model, the global minimum
point of which will determine the quantities $\sigma$ and $\Delta$,
is given by $S_{\rm eff}=-V_{\rm eff}\int d^4x \sqrt{-g}$, where
\begin{equation}
    V_{\rm
    eff}=\frac{3\sigma^2}{2G_1}+\frac{3\Delta\Delta^{*}}{G_2}+\tilde V_{\rm eff};\quad
   \tilde V_{\rm eff}=-\frac{ S_q}{v},\quad v=\int d^4x\sqrt{-g}.
\end{equation}

\section{Thermodynamic potential}

In this work we will consider the ultrastatic spacetime $R\otimes
H^3$ with constant negative curvature. The metric is given by
\begin{equation}
ds^2=dt^2-a^2(d\theta^2+\sinh^2\theta d\Omega_2),
\end{equation}
where $a$ is the radius of the hyperboloid which is related to the
scalar curvature by the relation $R=-\frac6{a^2}$, and $d\Omega_2$
is the metric on the two dimensional unit sphere.

Let us next introduce the one-particle Hamiltonian
$\hat{H}=-i\vec{\alpha} \vec{\nabla}+ \sigma\gamma^0$, where
$\vec{\alpha}=\gamma^0\vec{\gamma}$. Using this operator the quark
contribution to the effective action can be written in the following
form (for more details see~\cite{E_Zh_Tyu}):
\begin{equation}
\label{quark} S_q =
-\frac{i}{2}\left\{\ln\Det\Bigl[\hat{H}^2-(\hat{p}_0+\mu)^2\Bigr]+
    2\ln\Det\left[4|\Delta|^2+(\hat{H}-\mu)^2-\hat{p}_0^2\right]
        \right\},
\end{equation}
where $\hat{p}_0=i\partial_0$, and we have summed over colors (the
Det-operator now does not include color space).

In order to calculate the effective action one needs to solve the
equation for eigenfunctions of the Hamiltonian $\hat{H}$ in
hyperbolic space
\begin{equation}
(-i\vec{\alpha} \vec{\nabla}+
\sigma\gamma^0)\Psi_{\lambda}=\lambda\Psi_{\lambda}.
\end{equation}

Decomposing $\Psi_{\lambda}$ into upper $\psi_{\lambda,1}$ and lower
$\psi_{\lambda,2}$ components and further separating variables by
setting $\psi_{1,2}=f_{1,2}(\theta)\chi_{1,2}$, the solutions can be
found in terms of hypergeometric functions (for details see e.g.
~\cite{bytsenko, gorbar} and ~\cite{Weinberg, Camporesi}).

Let us now consider the diagonal element:
\begin{equation}
\tr \bra{\vec{x}} \ln\left[ \hat{H}^2-(\hat{p}_0+\mu)^2\right]
\ket{\vec{x}} = \int_0^\infty d \rho   \sum_{\xi=\pm}
\sum_{slm}\ln\left[ (\xi\lambda(\rho))^2-(\hat{p}_0+\mu)^2\right]
\psi_{\xi \rho l m }^{(s)}(\vec{x})^{\dag}\psi_{\xi \rho l m
}^{(s)}(\vec{x}),
\end{equation}
where $l$, $m$, $\rho=a\sqrt{\lambda^2-\sigma^2}$ and $s$ are the
quantum numbers which characterize the spinor $\chi$. Here $\xi$ is
the sign of the energy and we will now take $\lambda(\rho)>0$.
Notice that the last sum does not depend on the point $\vec{x}$ due
to the homogeneity of
the space $H^N$. Therefore it can be evaluated at
the origin $\theta=0$, when only the $l=0$ term survives
\cite{Camporesi}
(for generality, we quote below the formulae referring to the
N-dimensional space):
\begin{equation}
\sum_{slm} \psi_{\xi \rho l m }^{(s)}(0)^{\dag}\psi_{\xi \rho l m
}^{(s)}(0) = 2^{\frac{N-1}2} \mu_N(\rho), \quad \mu_N(\rho)\equiv
\frac1{a^N}\frac{\Gamma(N/2)2^{N-3}}{\pi^{N/2+1}}
|C_{l=0}(\rho)|^{-2}, \label{wfunctions}
\end{equation}
where
\begin{equation}
C_l(\rho)=\frac{2^{N-2}}{\sqrt{\pi}}\frac{\Gamma(l+N/2)\Gamma(i\rho+1/2)}{\Gamma(i\rho+l+N/2)}.
\end{equation}
Notice that the definition of the measure $\mu_N(\rho)$ (the density
of states) coincides with that used in \cite{bytsenko}.

Thus we can calculate the contribution from the first term in
(\ref{quark}):
\begin{eqnarray}
\ln\Det\Bigl[\hat{H}^2-(\hat{p}_0+\mu)^2\Bigr]&=&\Tr \ln
\Bigl[\hat{H}^2-(\hat{p}_0+\mu)^2\Bigr] = N_f \int d^N x \, dt \;
\tr \bra{\vec{x},t} \ln [\hat{H}^2-(\hat{p}_0+\mu)^2]
\ket{\vec{x},t} \nonumber\\
&=& v N_f 2^{\frac{N-1}2}\int \frac{dp_0}{2\pi}
\sum_{\xi=\pm}\int_0^\infty d\rho\, \mu_N(\rho) \ln\left[
(\xi\lambda(\rho))^2-(p_0+\mu)^2\right],
\end{eqnarray}
where $v$ is the spacetime volume of $R\otimes H^{N}$,
and $N_f=2$ is the number of flavours.

The second term gives:
\begin{equation}
\ln\Det\left[4|\Delta|^2+(\hat{H}-\mu)^2-\hat{p}_0^2\right] = v N_f
2^{\frac{N-1}2}\int \frac{dp_0}{2\pi} \sum_{\xi=\pm}\int_0^\infty
d\rho\, \mu_N(\rho)\ln\left[
4|\Delta|^2+(\lambda(\rho)+\xi\mu)^2-p_0^2 \right].
\end{equation}
Thus the quark contribution to the effective potential reads
\begin{equation}
    \label{V}
 \tilde V_{\rm eff}=-\frac{ S_q}{v}=\frac i2 N_f 2^{\frac{N-1}2}\int\frac{dp_0}{2\pi}
\sum_{\xi=\pm}\int_0^\infty d\rho\, \mu_N(\rho)
\Bigl\{\ln\left[(\lambda(\rho)+\xi\mu)^2-p_0^2\right]
          +2\ln\left[4|\Delta|^2+(\lambda(\rho)+\xi\mu)^2-p_0^2\right]\Bigl\}.
\end{equation}

In the case of finite temperature $T=1/\beta>0$, the following
substitutions should be made:
$$
    \int\frac{dp_0}{2\pi}(\cdots)\rightarrow \frac i{\beta} \sum_n(\cdots),\quad
    p_0\rightarrow i\omega_n,\quad
    \omega_n=\frac{2\pi}{\beta}\left(n+\frac12\right),
    \quad n=0,\pm1\pm2,\ldots,
$$
where $\omega_n$ is the Matsubara frequency. Then the quark
contribution to the effective potential (\ref{V}) becomes the
thermodynamic potential $\Omega_q$:
\begin{equation}
\Omega_q = - \frac{ 2^{\frac{N-1}2}N_f}
{2\beta}\sum_{n=-\infty}^{+\infty} \sum_{\xi=\pm}\int_0^\infty d
\rho \, \mu_N(\rho)
\Bigl\{\ln\left[(\lambda(\rho)+\xi\mu)^2+\omega_n^2\right]
          +2\ln\left[4|\Delta|^2+(\lambda(\rho)+\xi\mu)^2+\omega_n^2\right]\Bigl\}.
\end{equation}
Summing over the Matsubara frequencies we obtain the thermodynamic
potential:
\begin{eqnarray}
\Omega(\sigma,\Delta)&=&N_c\left(\frac{\sigma^2}{2G_1}+\frac{|\Delta|^2}{G_2}\right)
      -2^{\frac{N-1}2}N_f(N_c-2)\int_0^{\infty} d\rho \, \mu_N(\rho)\left\{ \lambda(\rho)+T\ln\left(1+e^{-\beta(\lambda(\rho)-\mu)}\right)+
       T\ln\left(1+e^{-\beta(\lambda(\rho)+\mu)}\right)\right\} \nonumber\\
    &&  -2^{\frac{N-1}2}N_f\int_0^{\infty} d\rho \, \mu_N(\rho) \biggl\{\sqrt{(\lambda(\rho)-\mu)^2+4|\Delta|^2}
            +\sqrt{(\lambda(\rho)+\mu)^2+4|\Delta|^2}+\\
    &&          +2T\ln\left(1+e^{-\beta\sqrt{(\lambda(\rho)-\mu)^2+4|\Delta|^2}}\right)+
                2T\ln\left(1+e^{-\beta\sqrt{(\lambda(\rho)+\mu)^2+4|\Delta|^2}}\right)\biggl\}.\nonumber
\end{eqnarray}
The spectrum of the Dirac operator $\lambda$ depends only on one
dimensionless parameter $\rho$. Instead of $\rho$ one can introduce
the quantity with the dimension of momentum $p=\rho/a$. Then the
spectrum of the Dirac operator may be written as
\begin{equation}
\lambda(p)=\sqrt{p^2+\sigma^2} \equiv E_p.
\label{energy}
\end{equation}
In the case of the 3D space, $N=3$, the density of states is
\begin{equation}
\mu_3(\rho)=\frac{\rho^2+\frac14}{2\pi^2a^3}. \label{density}
\end{equation}

Thus the thermodynamic potential in the $R\otimes H^3$ spacetime
becomes\footnote{ Note that after introducing the quantity with
dimension of momentum $p=\rho/a$ the spectrum of the Dirac operator
(\ref{energy}) formally coincides with the usual dispersion relation
in flat Minkowski spacetime and does not depend on curvature. At the
same time, the density of states (\ref{density}) in the hyperbolic
space differs from the usual measure of integration over momentum in
Minkowski spacetime by an additional term depending on the curvature
(in fact, this is the only curvature dependent part of the
thermodynamic potential). To obtain the correct measure of
integration over the continuum spectrum one needs to use the
properly normalized eigenfunctions in (\ref{wfunctions}).}:
\begin{eqnarray}
\Omega(\sigma,\Delta)=3\left(\frac{\sigma^2}{2G_1}+\frac{|\Delta|^2}{G_2}\right)
      &-&\frac{2}{\pi^2}\int_0^{\infty} d p ~(p^2+\frac1{4a^2})\big\{ E_p+T\ln\left(1+e^{-\beta(E_p-\mu)}\right)+
       T\ln\left(1+e^{-\beta(E_p+\mu)}\right)\nonumber\\
&+& \sqrt{(E_p-\mu)^2+4|\Delta|^2}+\sqrt{(E_p+\mu)^2+4|\Delta|^2} \\
&+&2T\ln\left(1+e^{-\beta\sqrt{(E_p-\mu)^2+4|\Delta|^2}}\right)+
                2T\ln\left(1+e^{-\beta\sqrt{(E_p+\mu)^2+4|\Delta|^2}}\right)\big\}.\nonumber
\end{eqnarray}

It should be noted that the thermodynamic potential is divergent at
large $p$. Therefore we must use some regularization procedure.
Since we interpret $p$ as the module of the momentum, the easiest
way to regularize the divergent integral is to introduce the
momentum cutoff, $p\leq \Lambda$.
Then the regularized potential looks as follows:
\begin{eqnarray}
\label{potential}
\Omega^{\mathrm{reg}}
(\sigma,\Delta)=3\left(\frac{\sigma^2}{2G_1}+\frac{|\Delta|^2}{G_2}\right)
      &-&\frac{2}{\pi^2}\int_0^{\Lambda} d p ~(p^2+\frac{|R|}{24})\big\{ E_p+T\ln\left(1+e^{-\beta(E_p-\mu)}\right)+
       T\ln\left(1+e^{-\beta(E_p+\mu)}\right)\nonumber\\
&+& \sqrt{(E_p-\mu)^2+4|\Delta|^2}+\sqrt{(E_p+\mu)^2+4|\Delta|^2} \\
&+&2T\ln\left(1+e^{-\beta\sqrt{(E_p-\mu)^2+4|\Delta|^2}}\right)+
                2T\ln\left(1+e^{-\beta\sqrt{(E_p+\mu)^2+4|\Delta|^2}}\right)\big\},\nonumber
\end{eqnarray}
where we used the expression for the scalar curvature
$R=-\frac6{a^2}$. This formula consists of two parts arising from
the two terms in the first round bracket of the integrand: the first
part is the same as in (3+1)D Minkowski spacetime, while the second
one, which is linear in curvature, corresponds to the contribution
from (1+1)D spacetime. Hence, when the contribution of the second
term dominates over the first one, dimensional reduction by two
units takes place.

The values of condensates $\sigma$ and $|\Delta|$ correspond to the
point of global minimum of the regularized thermodynamic potential
and are determined as solutions of the gap equations
\begin{equation}
\frac{\partial \Omega^{\rm reg}}{\partial \sigma} =0 , \qquad
\frac{\partial \Omega^{\rm reg}}{\partial |\Delta|} =0.
\end{equation}

In the following sections we will consider the behavior of the
condensates as functions of
curvature, temperature and chemical potential.

\section{Analytical solutions}

\subsection{Chiral condensate}

Let us first consider the case when chiral symmetry is broken while
the color symmetry remains unbroken ($\sigma\neq0$ and $\Delta=0$).
Since the quark condensate appears even in the vacuum, we put for
simplicity $\mu=0$ and $T=0$.

Then one can obtain the following expression for the effective
potential
\begin{equation}
V_{\rm eff}(\sigma) = \frac{\Lambda^4}{\pi^2}v_0(x), \quad
v_0(x)=\frac{3x^2}{2g}-\frac{3}{4}F(x)-\frac{|r|}{8}G(x), \quad
x=\frac{\sigma}{\Lambda},  \, g=\frac{\Lambda^2}{\pi^2}G_1,\,
r=\frac R{\Lambda^2},
\end{equation}
where
\begin{equation}\label{FG}
F(x)=(2+x^2)\sqrt{1+x^2}-x^4\ln{\frac{1+\sqrt{1+x^2}}{x}}, \quad
G(x)=\sqrt{1+x^2}+x^2\ln{\frac{1+\sqrt{1+x^2}}{x}}.
\end{equation}
The gap equation for the condensate $\sigma$ reads
\begin{equation}
\label{gap}
\frac1g=\sqrt{1+x^2}+(\frac{|r|}{12}-x^2)\ln{\frac{1+\sqrt{1+x^2}}{x}}.
\end{equation}
An analytical solution of this equation can be obtained only for
small $x\ll1$ ($\sigma\ll\Lambda$). Expanding the right hand side of
(\ref{gap}) in $x$ we have
\begin{equation}
\label{approx}
\left(\frac1g-1\right)=(\frac{|r|}{12}-x^2)\ln{\frac{2}{x}}.
\end{equation}

We will consider three different cases: a) subcritical $g$,
$g<g_c=1$, where $g_c$ is the critical constant in flat four
dimensional spacetime; b) near critical $g$, when $g\to g_c-0$ and
c) overcritical $g$, $g>1$.

In the case of subcritical $g$, a nontrivial solution of the gap
equation (\ref{approx}) exists only if $x^2<\frac{|r|}{12}$. In the
strong curvature limit $\sigma^2\ll\frac{|R|}{12}$, the chiral
condensate is given by
\begin{equation}
\label{cond} \sigma_0=2\Lambda \exp{\left[
-\frac{12}{|r|}\left(\frac1g-1\right) \right]}=2\Lambda \exp{\left[
-\frac{12\pi^2(1-g)}{|R|G_1}
 \right]}.
\end{equation}
One can distinguish two sub-cases in which the last expression is
consistent with the above made assumptions: \\ i) $g<1$, $r\ll1$ or
ii) $g\ll1$, $r \sim 1$.
The case ii) shows that the gap equation has a nontrivial solution
even at arbitrary weak coupling constant.

The expression (\ref{cond}) looks very similar to the chiral
condensate in the two-dimensional Gross-Neveu model. After excluding
the coupling constant from the effective potential by using the gap
equation (\ref{approx}), we obtain
\begin{equation}
V_{\rm eff}(\sigma) = \frac{|R|\sigma^2}{16\pi^2}\left(\ln
\frac{\sigma^2}{\sigma^2_0}-1\right),
\end{equation}
and this is indeed (up to a dimensional factor) the effective
potential of the Gross-Neveu model. Hence, we conclude that in the
case of subcritical coupling the
strong gravitational field
of hyperbolic space leads to the effective dimensional reduction
from (3+1) to (1+1) (see also \cite{gorbar}).

One should also note that the non-analytical dependence of the
chiral condensate on curvature in the exponent of (\ref{cond}) looks
quite similar to that in a magnetic field
\begin{equation}
\sigma_0=\sqrt{\frac{|eB|}{\pi}}\exp{\left[-\frac{2\pi^2(1-g)}{|eB|G_1}\right]},
\end{equation}
but the pre-exponential factors differ (see for example the review
\cite{gusynin}). This fact demonstrates that in the case of
hyperbolic space the negative curvature plays an analogous
catalyzing role as the magnetic field does in flat space, where it
gives rise to the catalysis of $\chi$SB even at arbitrary weak
attraction between quarks.

In the near critical regime $g\to g_c -0$ the chiral condensate is
just the constant
\begin{equation}
\sigma_0=\sqrt{\frac{|R|}{12}},
\end{equation}
where the curvature must be small $\frac{|R|}{12}\ll\Lambda^2$.

In the overcritical regime $g>1$, a nontrivial solution of
(\ref{approx}) exists only if $x^2 > \frac{|r|}{12}$. In the
weak curvature limit
$\frac{|R|}{12}\ll m^2$ we obtain
\begin{equation}
\sigma_0=m\left(1+\frac{|R|}{24m^2}+O\left(\frac{R^2}{m^4}\right)
\right), \label{correction}
\end{equation}
where $m$ is the solution of the gap equation at $R=0$. It is seen
that in this case the curvature leads to small analytical
corrections to the flat-space value of chiral condensate.

Next, let us consider the influence of finite temperature on the
behavior of the chiral condensate. The temperature dependent
contribution to the thermodynamic potential looks as follows
\begin{equation*}
\Omega_T(\sigma)=-\frac{12}{\pi^2}T \int_0^{\infty}dp
\left(p^2+\frac{|R|}{24} \right)\ln{\left(1+e^{-\beta E_p} \right)}.
\end{equation*}
Expanding the logarithm into a series and performing the integration
over momentum, we obtain
\begin{equation*}
\Omega_T(\sigma)=\frac{12}{\pi^2} T \sigma
\left(\frac{|R|}{24}\sum_{n=1}^{\infty}\frac{(-1)^n}{n}K_1(n\beta\sigma)+
\sigma^2\sum_{n=1}^{\infty}\frac{(-1)^n}{n}\frac{K_2(n\beta\sigma)}{n\beta\sigma}
\right),
\end{equation*}
where $K_{\nu}(x)$ is the Macdonald function (modified Bessel
function). Then the gap equation at finite temperature reads
\begin{equation}
\frac1g=\sqrt{1+x^2}+(\frac{|r|}{12}-x^2)\ln{\frac{1+\sqrt{1+x^2}}{x}}-\frac{|r|}{12}I_1
-x^2(I_3-I_1),
\end{equation}
where
\begin{equation*}
I_1(\beta\sigma)=-2\sum_{n=1}^{\infty}(-1)^n K_0(n\beta\sigma),
\quad I_3(\beta\sigma)=-2\sum_{n=1}^{\infty}(-1)^n
K_2(n\beta\sigma).
\end{equation*}
Here we consider only the most interesting case of subcritical
coupling
and strong gravitational field.
Thus, as previously at $g<1$ and $\sigma^2\ll\frac{|R|}{12}$, we
obtain for the chiral condensate
\begin{equation}
\sigma_0(T)=2\Lambda \exp{\left[ -\frac{12\pi^2(1-g)}{|R|G_1} -
I_1(\beta\sigma_0(T))
 \right]}=\sigma_0(0)\exp{\left[-I_1(\beta\sigma_0(T))\right]}.
\label{temperature}
\end{equation}
The function $I_1(x)$ is positive and monotonically decreases which
means that temperature leads to the restoration of broken symmetry.
The critical temperature $T_c$ is defined by the condition
$\sigma_0(T_c)=0$ which gives the BCS-like relation
\begin{equation}
\label{Tc}
 T_c=\pi^{-1}e^{C}\sigma_0(0)\simeq 0,57 \sigma_0(0),
\end{equation}
where $\sigma_0(0)$ is given by (\ref{cond}).
It should be mentioned
that the temperature dependence of the chiral condensate
(\ref{temperature}) is the same as
in a constant chromomagnetic field \cite{toki}.

This brief analysis shows that in the case of subcritical coupling
the negative curvature leads to the catalysis of $\chi$SB while in
the overcritical regime it leads to the enhancement of the chiral
condensate. One might expect the same behaviour in the case of a
color condensate. The case of overcritical coupling will also be
considered in the next section using numerical methods.

\subsection{Mixed phase}

Let us now consider a more general situation, when both condensates
can take nonzero values.
What concerns color symmetry breaking in flat spacetime, it is well
known that  diquark pairing  and color superconductivity arise (for
large quark number densities and, correspondingly, at large chemical
potential) due to an instability of the Fermi surface so that
the color superconducting state is energetically more preferable.
For studying the influence of gravity on the formation of
a color condensate in its most ``pure form'',
we find it convenient, contrary to the
flat case, in the following to take $\mu=0$. Then the effective potential at
zero temperature can be written in the form
\begin{equation}
\label{eff} V_{\rm eff}(\sigma,\Delta) =
\frac{\Lambda^4}{\pi^2}v_0(x,y), \quad v_0(x)=\frac{3A}{2}x^2+By^2
-\frac{1}{4}\left(F(x)+2F(z)\right)-\frac{|r|}{24}\left(G(x)+2G(z)\right),
\end{equation}
where
\begin{equation}
\label{const}
    x=\frac{\sigma}{\Lambda}, \quad y=\frac{2|\Delta|}{\Lambda},\quad z=\frac{m_{\ast}}{\Lambda}=\sqrt{x^2+y^2}, \quad A={1\over g}=\frac{\pi^2}{\Lambda^2 G_1}, \quad
 B=\frac{3\pi^2}{4\Lambda^2 G_2}, \quad
r=\frac R{\Lambda^2},
\end{equation}
and the functions $F$ and $G$ are the same as in (\ref{FG}). The
nontrivial solutions for condensates satisfy the following gap
equations obtained from (\ref{eff})
\begin{eqnarray}
    3A&=&H(x)+2H(z)+\frac{|r|}{12}(K(x)+2K(z)),  \label{gap1} \\
     B&=&H(z)+\frac{|r|}{12}K(z), \label{gap2}
\end{eqnarray}
where
\begin{equation}
    H(x)=\sqrt{1+x^2}-x^2\ln{\frac{1+\sqrt{1+x^2}}{x}}, \quad
    K(x)=\ln{\frac{1+\sqrt{1+x^2}}{x}}.
\end{equation}
Substituting (\ref{gap2}) in (\ref{gap1}), we obtain a separate
equation for the chiral condensate
\begin{equation}\label{gap3}
    3A-2B=H(x)+\frac{|r|}{12}K(x),
\end{equation}
which formally coincides with (\ref{gap}) in the previous Subsection
\textbf{A}, when $1/g$ becomes replaced by $(3A-2B)$. As in the
previous case, we will solve equations (\ref{gap3}) and (\ref{gap2})
in the limit of small condensates $x\ll1$ and $y\ll1$ and consider
only the most interesting case of strong curvature $x^2\ll
\frac{|r|}{12}$, $y^2 \ll \frac{|r|}{12}$ and subcritical couplings
$A>1$, $B>1$, where the effect of gravitational catalysis is more
clearly seen. In this limit we simply have $H(x)=1$ and
$K(x)=\ln{(2/x)}$. The solutions of the gap equations are as follows
\begin{eqnarray}
 \sigma^2_0=4\Lambda^2\exp{\left[-\frac{24}{|r|}(3A-2B-1) \right]}, \label{sigma0} \\
 m^2_{\ast0}=\sigma^2_0+4|\Delta_0|^2 = 4\Lambda^2\exp{\left[-\frac{24}{|r|}(B-1)
 \right]} \label{m0},
\end{eqnarray}
where the coupling constants must satisfy the inequalities: $3A-2B>1$ and
$B>1$. From this, the color condensate follows
\begin{equation}
\label{delta}
    4|\Delta_0|^2=m^2_{\ast0}\left\{1-\exp{\left[-\frac{24}{|r|}(3A-3B)\right]}\right\},
\end{equation}
which exists only if $A>B$ (i.e. $G_2>\frac{3}{4}G_1$). Hence, both
condensates may exist simultaneously in the region $A>B>1$ (in this
region the inequality $3A-2B=A+2(A-B)>1$ holds automatically).

Excluding the coupling constants from the effective potential one
obtains
\begin{equation}
    V_{\rm eff}(\sigma, \Delta)=\frac{|R|}{48\pi^2}\left[ \sigma^2\left(\ln{\frac{\sigma^2}{\sigma^2_0}}-1\right)+2m_{\ast}^2\left(\ln{\frac{m_{\ast}^2}{m_{\ast0}^2}}-1\right) \right]
\end{equation}

The other possible stationary point of the effective potential is
$\sigma=0$ and $\Delta\neq0$. This type of solution may be obtained
from the previous one by letting $\sigma_0^2=0$ in (\ref{m0})
so that
the combined condensate $m_{\ast0}^2$ reduces to
the tilded expression
$4|\widetilde{\Delta}_0|^2$.

The phase structure of the model is defined by the
global minimum of the effective potential. There are four types of
stationary points of the effective potential: $(0,0)$,
$(\widetilde{\sigma}_0,0)$, $(0,\widetilde{\Delta}_0)$ and
$(\sigma_0,\Delta_0)$. Let us denote the corresponding values of the
effective potential at
these
points by $v_1$, $v_2$, $v_3$ and $v_4$.
The effective potential is normalized in such a way that $v_1=0$.
The other values are
\begin{equation}
    v_4=-\frac{|R|}{48}\left[\sigma^2_0+2m^2_{\ast0}\right], \quad
    v_3=-\frac{|R|}{6}|\widetilde{\Delta}_0|^2, \quad
    v_2=-\frac{|R|}{16}\widetilde{\sigma}^2_0,
\end{equation}
where $\widetilde{\sigma}_0$ is given by (\ref{cond}). First of all
we see that the minimum $v_1=0$ of the symmetry case is higher than
the other minima. This means in contrast to the flat case that for
subcritical couplings the symmetric phase in hyperbolic space is
unstable under the formation of different condensates and symmetry
breaking. Since $m_{\ast0}^2=4|\widetilde{\Delta}_0|^2$, it is
easily seen that $v_4<v_3$ in the region $A>B>1$. The mixed phase is
the true vacuum of our model if $v_4<v_2$ or
$\sigma^2_0+2m^2_{\ast0}>3\widetilde{\sigma}^2_0$. Dividing both
sides of the last inequality by $\widetilde{\sigma}^2_0$ and using
(\ref{cond}), (\ref{sigma0}) and (\ref{m0}) we obtain the following
condition
\begin{equation*}
    \exp{\left[\frac{24}{|r|}2(B-A) \right]}+2\exp{\left[\frac{24}{|r|}(A-B)
 \right]} > 3.
\end{equation*}
Introducing a new variable $u=\exp{\left[\frac{24}{|r|}(B-A)
\right]}$, $0<u<1$ for $A>B>1$, we see that the above condition is
just the simple cubic inequality $u^3+2>3u$, which is automatically
satisfied for $0<u<1$. Thus, the condition $A>B>1$ is sufficient for
the mixed phase to be the true vacuum state of our model.

In the opposite case $B\geq A>1$ the chiral and color condensates
cannot exist simultaneously and we need to compare $v_2$ and $v_3$.
The inequality $v_3<v_2$ leads to $A-B>\frac{|r|}{24}\ln\frac32$
which contradicts our previous assumption. Therefore in the region
$B\geq A>1$ only the phase with broken chiral symmetry occurs, and
the chiral condensate is described by (\ref{cond}).

Let us also briefly consider the influence of finite temperature on
the mixed phase. Proceeding in the same manner as in Subsection
\textbf{A}, we obtain the following temperature dependence of
condensates
\begin{eqnarray}
    \sigma^2_0(T)&=&\sigma^2_0(0)\exp{\left[ -2I_1(\beta
    \sigma_0(T))\right]}, \label{sigma} \\
    m^2_{\ast0}(T)&=&m^2_{\ast0}(0)\exp{\left[- 2I_1(\beta
    m_{\ast0}(T))\right]}, \label{m}
\end{eqnarray}
where $\sigma^2_0(0)$ and $m^2_{\ast0}(0)$ are given by
(\ref{sigma0}) and (\ref{m0}), respectively. In analogy with
(\ref{Tc}), the critical temperatures for chiral and color
condensates are
\begin{equation}
    T^{\sigma}_c=\pi^{-1}e^{C}\sigma_0(0), \qquad
    T^{\Delta}_c=\pi^{-1}e^{C}m_{\ast0}(0).
\label{sigmadelta}
\end{equation}
Since the inclusion of temperature results in a smooth diminution of
condensates, the relation between couplings which determines the
phase structure of the model should not change. Again both
condensates may exist simultaneously if $A>B$ which implies the
relation $T^{\Delta}_c>T^{\sigma}_c$. Therefore at $T<T^{\sigma}_c$
the mixed phase is realized with chiral and color condensate of the
form
\begin{eqnarray}
    |\Delta_0(T)|^2=\frac14\left\{m^2_{\ast0}(T)-\sigma^2_0(T)\right\}.
\end{eqnarray}
Obviously, at $T^{\sigma}_c<T<T^{\Delta}_c$, chiral symmetry is
restored, while the color symmetry remains broken
\begin{equation}
|\Delta_0(T)|^2=|\widetilde{\Delta}_0(T)|^2=\frac{m^2_{\ast0}(T)}{4}.
\end{equation}
Finally, at $T>T^{\Delta}_c$ both symmetries become restored.

\begin{figure}[h]
\noindent
 \centering{
  $
\begin{array}{cc}
 \epsfig{file=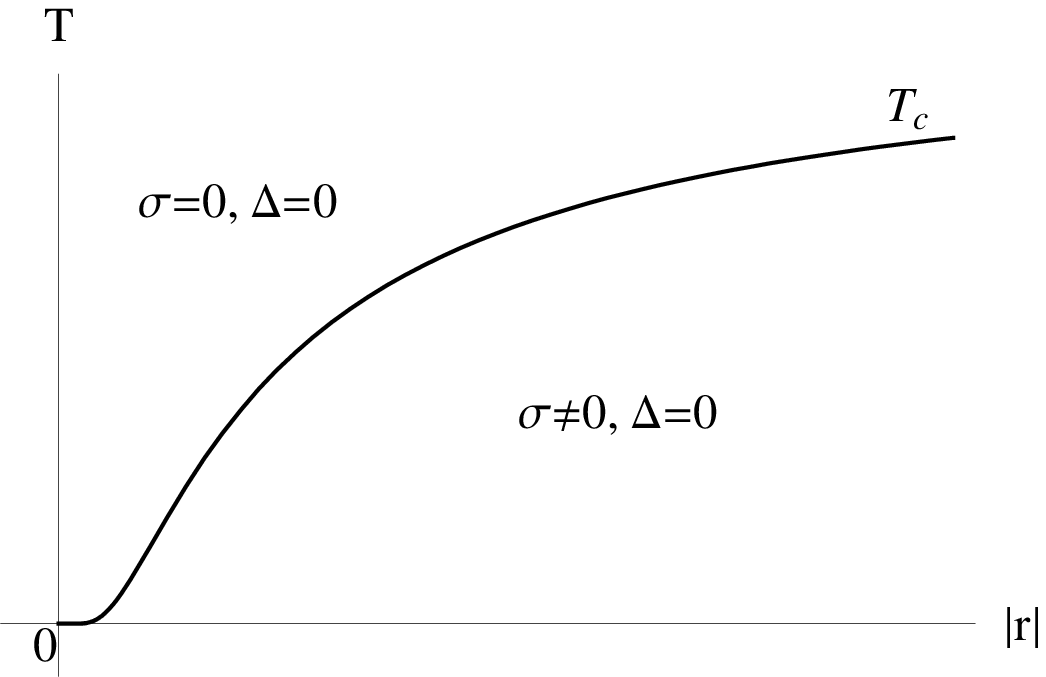,width=8.5cm}&
 \epsfig{file=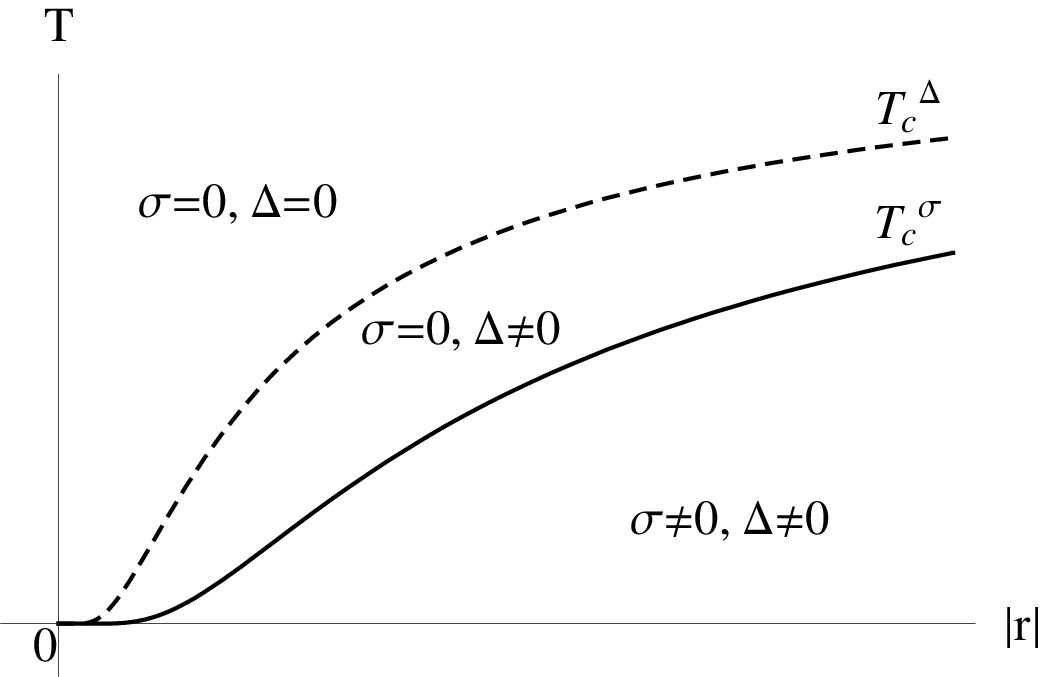,width=8.5cm}
  \end{array}
    $
}\caption{
The phase portraits of the system
showing critical
curves as functions of $|r|$ for fixed values of
couplings $A$ and $B$: for $A\leq B$ (left) and for $A>B$ (right).}
\label{f0}
\end{figure}

The phase portraits of the system are
presented in Fig.\ref{f0} for
fixed  values of coupling constants A and B, for $A\leq B$ (left
picture) and for $A>B$ (right picture).   
The critical curves which separate different phases are described by
corresponding formulas (\ref{Tc}), (\ref{sigmadelta}) for critical
temperatures.

\begin{figure}[h]
\epsfig{file=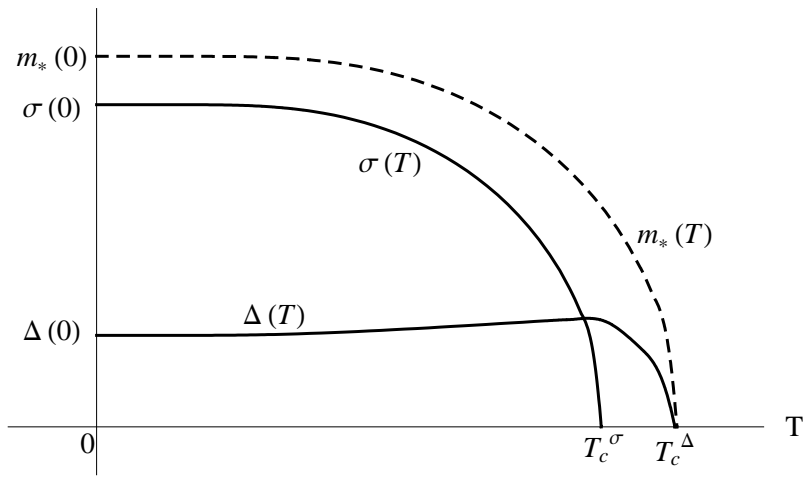} \caption{
Condensates
$\sigma_0$, $\Delta_0$
and $m_{\ast0}$ as functions of temperature inside the mixed phase.
(Henceforth we will omit the subscript 0 of condensates on the plots.)}
\label{f}
\end{figure}

In fact the solutions (\ref{sigma}), (\ref{m}) of the gap equations
give only an implicit dependence of the condensates $\sigma_0(T)$
and $m_{\ast0}(T)$ on temperature. Due to the complexity of the
function $I_1(x)$, no analytical solutions for condensates can be
found. However, equations (\ref{sigma}), (\ref{m}) can be easily
solved by using numerical methods, for example, an iterative
procedure. When the temperature is fixed, the result of such
procedure will depend only on the first step of iteration. The most
obvious choice is to take the condensates at zero temperature as a
first approximation. The schematic picture of condensates inside the
mixed phase is shown in Fig.\ref{f} for fixed values of
$\sigma_0(0)$ and $\Delta_0(0)$.
It is interesting to note that the condensates $\sigma_0$ and
$m_{\ast0}$ have the usual form of the BCS theory of
superconductivity and decrease with temperature, while the color
condensate $\Delta_0$ slightly grows at $T \lesssim T_c^{\sigma}$
and has a maximum around $T_c^{\sigma}$. We have not specified the
values of condensates on these curves since we have no
phenomenologically motivated choice of coupling constants and
curvature. Fig.\ref{f} demonstrates only qualitatively the behaviour
of condensates as functions of the temperature at fixed curvature
$|r|$ and couplings $A$ and $B$. Moreover, since the condensates at
nonzero temperature depend on the curvature only through the
condensates at zero temperature, the increasing curvature will
produce the appropriate stretching of curves in Fig.\ref{f}.

\section{Phase transitions}
In the previous section the phase structure of the model was
extensively studied in the regime of subcritical couplings by
formally considering arbitrary ratios of coupling constants. In
particular, in order to investigate the influence of curvature on
condensates in "pure form", the chemical potential was taken to be
zero. Now, in this section, we turn to the more general case of
phase transitions at finite chemical potential (quark number
density).

In the following we want to compare our results with the well
established case of flat spacetime. Since the role of a growing
temperature in the restoration of broken symmetries was already
demonstrated, our considerations will be restricted here to the case
of zero temperature. The investigation of the phase transitions at
finite chemical potential beyond the limit of small condensates
is difficult to perform analytically, and hence
numerical methods will now be used.

In order to be able to make comparison with the flat case, we
consider the following ``more physical'' relation of couplings taken
from the instanton-motivated NJL-model \cite{Berges} (see also
\cite{Ebert_Zhukovsky}):
\begin{equation}
    \label{G}
    G_2=\frac38G_1.
\end{equation}

Let us now fix $g=G_1\Lambda^2/\pi^2=1.4$ in such a way that the
chiral condensate at zero curvature, $R=0$, in the vacuum is equal
to the usual value of the constituent quark mass in flat space
($\sigma_0=350$ MeV). Here the cutoff parameter is taken to be
$\Lambda=600$ MeV.

In terms of dimensionless couplings A and B (see (\ref{const})),
relation (\ref{G}) corresponds to $B=2A$, and one might think that
it excludes the existence of a mixed phase in accordance with our
previous discussion. However, we should stress that the inequality
$B\geq A$, which guaranties the absence of a mixed phase, was
obtained only for subcritical couplings. Here we have fixed $g=1.4$
and thus $A=1/g<1$, and the overcritical regime is realized. The
difference between these cases is that in the subcritical regime the
main contribution to the values of condensates is given by the
two-dimensional part of the effective potential, while in the
overcritical regime, the condensates receive their contribution
mainly from the flat four-dimensional part. Therefore our previous
arguments are not applicable in this case.

It was observed earlier that in a flat four dimensional
spacetime the relation (\ref{G}) between coupling constants leads to the
absence of a mixed phase in a wide range of parameters
\cite{Berges}. In the previous section, we have found that in the
overcritical regime finite curvature gives only small corrections to
the flat-space value of chiral condensate (see (\ref{correction})).
Using numerical calculations, we will now analogously show that in a
wide range of the values of curvature its contribution to
condensates is small in comparison with their values in flat case.
It is clear that such corrections can not change the phase structure
of the model obtained in flat case in \cite{Berges}, and thus can
not lead to formation of new phases. Therefore, in what follows, we
assume that there is no mixed phase, where both condensates
simultaneously take nonzero values.

As is well-known, for increasing chemical potential there arises
diquark pairing, whereas the chiral condensate becomes suppressed.

\begin{figure}[h]
\noindent
 \centering{
  $
\begin{array}{cc}
 \epsfig{file=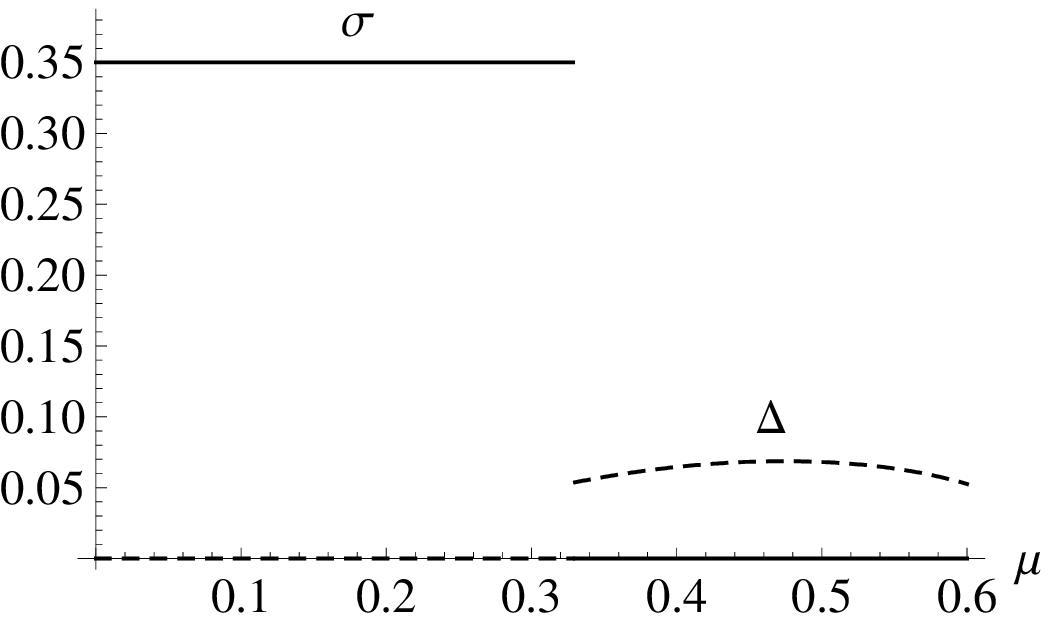,width=8.5cm}&
 \epsfig{file=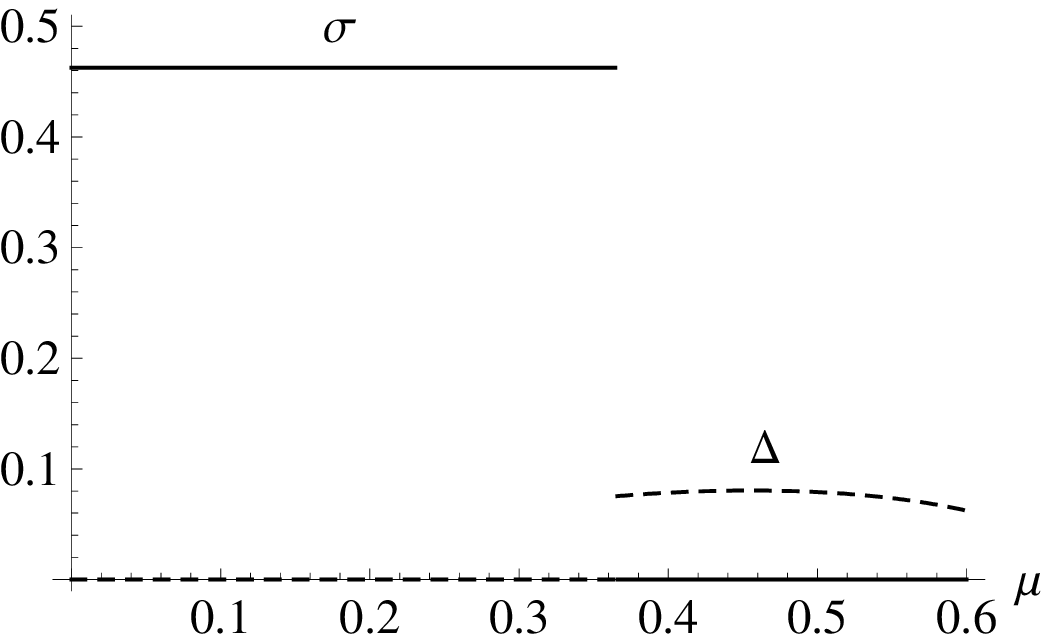,width=8.5cm}
  \end{array}
    $
}\caption{Condensates $\sigma$ and $\Delta$ as functions of $\mu$
(all quantities are given in units of GeV) at $r=0$ (left) and at
$|r|=1$ (right).} \label{f1}
\end{figure}

The corresponding behavior of the chiral and color condensates as
functions of the chemical potential at $r=0$ and $|r|=1$ is shown in
Fig.\ref{f1}.

As is seen from Fig.\ref{f1}, the critical chemical potential, at
which the phase transition takes place, in flat space, $r=0$, is
$\mu_c\approx330$ MeV, while at $|r|=1$ $\mu_c\approx366$ MeV. From
this we can conclude that the critical line $\mu_c(|r|)$ which
separates the two phases is a growing function of the curvature. The
numerical study of the number density, which is the first derivative
of the thermodynamical potential with respect to the chemical
potential, shows that it is discontinuous at $\mu_c$. Thus, an
increasing chemical potential leads to a first order phase
transition.

It should  also be mentioned that for the limiting case of zero
curvature our results for the value of the critical chemical
potential $\mu_c\approx 330$ MeV and the maximum value of the color
condensate $\Delta\approx 70$ MeV (see Fig.\ref{f1}) are in
agreement with the results obtained in \cite{blaschke} for the same
values of $\sigma_0$ and $\Lambda$ (note that  our value of the
color condensate $\Delta$ is by definition two times less than
condensate $\Delta$ in \cite{blaschke}).

\begin{figure}[h]
\noindent
 \centering{
  $
\begin{array}{cc}
 \epsfig{file=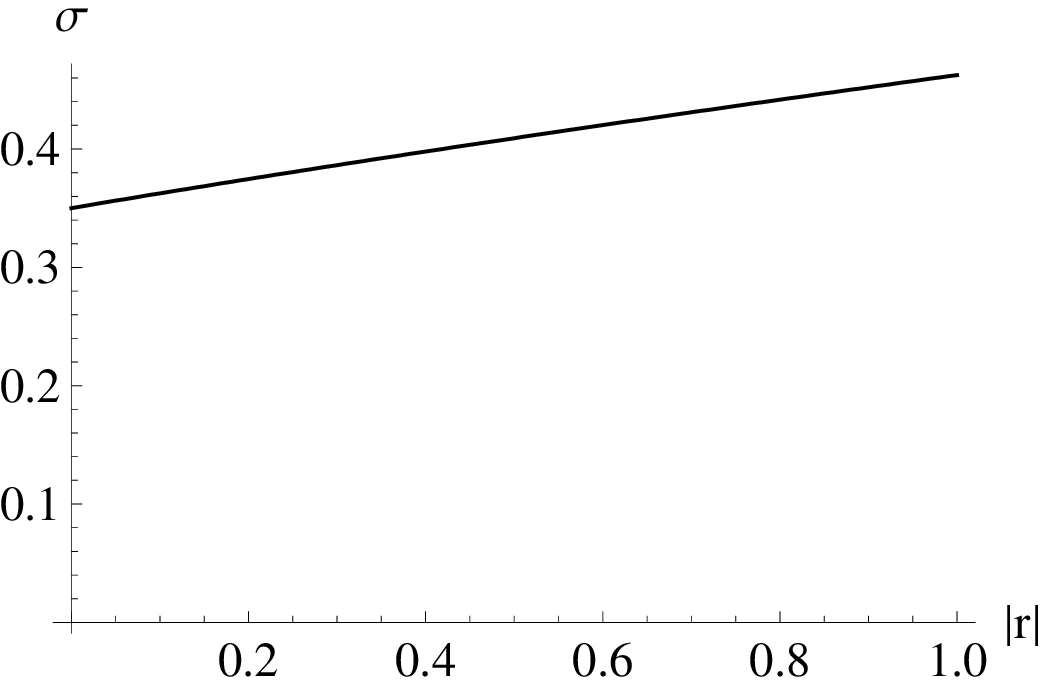,width=8.5cm}&
 \epsfig{file=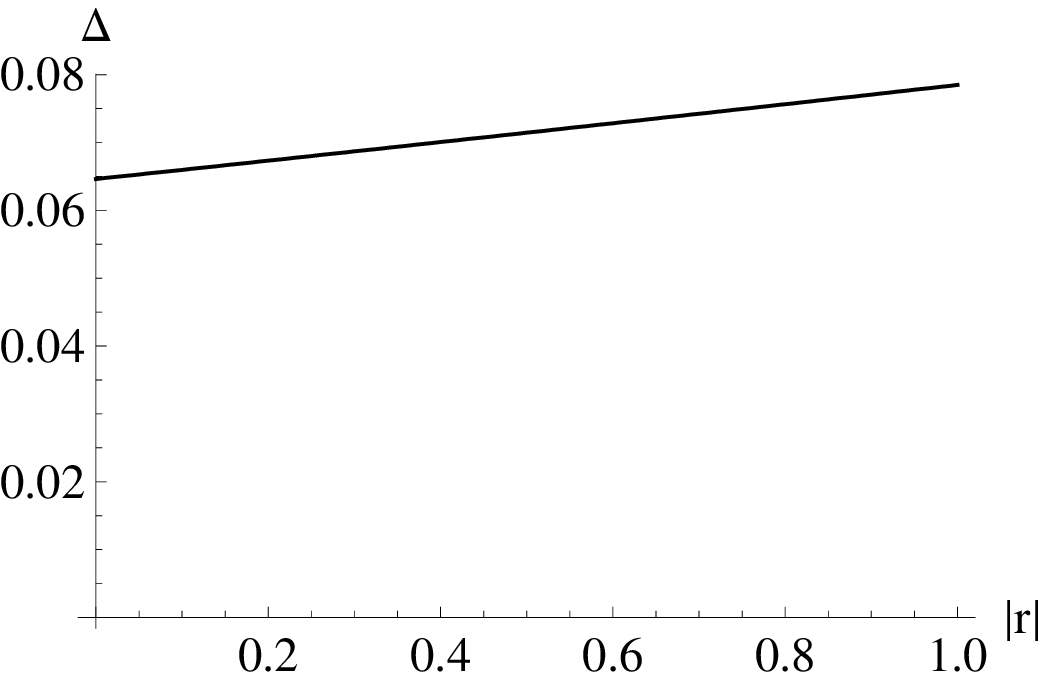,width=8.5cm}
  \end{array}
    $
}\caption{Condensates (in units of GeV) $\sigma$ at $\mu=200$ MeV
(left) and $\Delta$ at $\mu=400$ MeV (right) as functions of $|r|$.}
\label{f3}
\end{figure}

We can also examine the condensates $\sigma$ and $\Delta$ as
functions of the absolute value of curvature $|r|$. The behavior of
chiral and color condensates inside both phases is presented in
Fig.\ref{f3}.

It is clear that with growing curvature the values of condensates
slowly increase, which results in the enhancement of the symmetry
breaking effects. As we have already mentioned above, in a wide
range of curvature values the increment of the condensates is small
with respect to their values at $r=0$. Moreover, it is seen from
Fig.\ref{f3} that condensates grow with curvature almost linearly
which is in agreement with our previous analytical consideration
(see (\ref{correction})).

\begin{figure}[h]
\epsfig{file=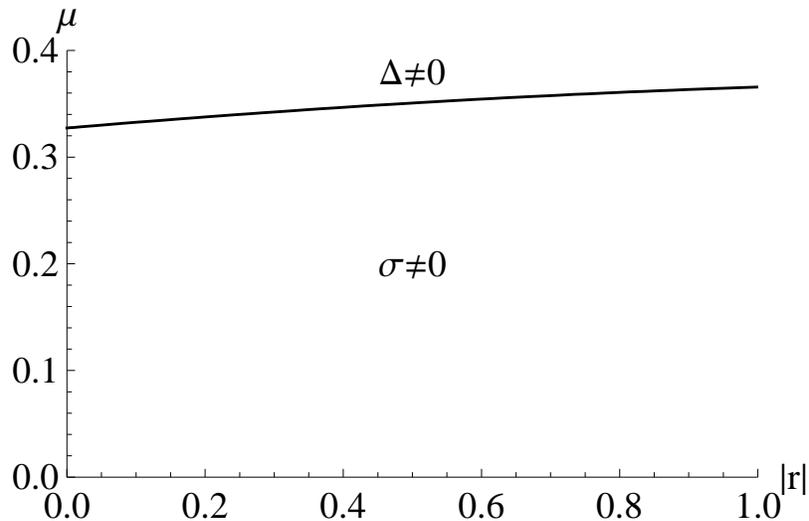} \caption{Phase portrait for the coupling
relation $G_2=\frac{3}{8}G_1$.} \label{f4}
\end{figure}

Finally, the $r-\mu$ - phase portrait of our system is shown in
Fig.\ref{f4}. As it was already mentioned, the critical chemical
potential, at which the first order phase transition takes place,
slightly grows with increasing curvature.

\section{Summary}

In the framework of an extended NJL model we have investigated
chiral and/or color symmetry breaking in dense quark matter under
the influence of negative curvature of hyperbolic space by employing
the thermodynamic (effective) potential of the system as a function
of chiral and diquark condensates, temperature, curvature and
chemical potential.

Two different regimes of dynamical symmetry breaking have been
studied. First, the regime of subcritical couplings, where the
values of coupling constants are lower than their critical values in
flat space, was considered. As is well known, in this situation in
four dimensions, no symmetry breaking takes place, and gap equations
have only trivial solutions. Unlike the flat case, in hyperbolic
space, symmetry may, however, be broken even for an arbitrary small
coupling constant. In the case of subcritical couplings
we obtained expressions for the chiral and color condensates that depend
non-analytically on curvature.
Secondly, the regime of overcritical coupling constants was investigated, when
the symmetry may be broken even in flat space. In this case it was
shown that curvature leads to
small analytical corrections which increase the flat-space values of
condensates and
thus enhances
the symmetry breaking effects.

It is interesting to note that in the subcritical regime of coupling
constants the
strong gravitational field
of hyperbolic space serves as a catalyzing factor similar to the
role of the magnetic field \cite{Klimenko,5,6} or chromomagnetic
fields \cite{KMV,ebzhu,toki,Ebert_Zhukovsky} in the effects of
dynamical symmetry breaking. As we have explicitly demonstrated,
the gravitational catalysis takes place for chiral and color
condensates. It is also worth mentioning that the effect of
gravitational catalysis is accompanied by
a lowering of dimensions of the system. In the regime of subcritical
couplings the solutions of gap equations look quite similar to those
for the 2D Gross-Neveu model. Therefore, we concluded that the
strong curvature of hyperbolic space leads to an effective dimensional
reduction by two units (see also \cite{gorbar}).

As we have already mentioned, in the subcritical regime the negative
curvature essentially changes the phase structure of the NJL model
found in \cite{Ebert_Khudyakov} making the symmetric phase unstable
under the formation of condensates,
while in the overcritical case one expects only minor modifications to the
phase structure obtained in flat space. Therefore we have
extensively studied the phase structure in the subcritical regime.
It is interesting to note that the overall phase structure depends
only on the ratio of the inverse (dimensionless) coupling constants
A and B, but not on the curvature. For subcritical couplings the
mixed phase is realized only if $A/B>1$, while if $A/B\leq1$ only
the chiral symmetry may be broken. The same critical ratio $A/B=1$
was found in the flat space in the framework of the random matrix
model \cite{vanderheyden}. It was argued by these authors that this
relation is a consequence of global symmetries of the model.

Moreover, the influence of finite temperature on phase structure was
investigated. The phase portraits of the system for different
relations between couplings were constructed. In particular, we have
shown that for any fixed value of curvature there exists a critical
temperature at which the phase transition takes place and symmetry
becomes restored.

Finally, using numerical calculations, we
have investigated the phase
transitions between $\chi$SB and CSC phases under the influence of
chemical potential and curvature in the regime of overcritical
couplings. It was demonstrated that similar to the flat case there
arises a diquark pairing for increasing chemical potential, while the chiral
condensate becomes suppressed. The phase portrait of the system at
zero temperature was also constructed, and it was shown that the
critical line $\mu_c(|r|)$, separating the two different phases, is
a growing function of curvature. The chiral and diquark condensates,
$\sigma$ and $\Delta$, acquire only small corrections due to
curvature increasing the flat-space values of condensates and this
leads to an enhancement of the symmetry breaking effects.

The results of this paper, although describing a model situation
with symmetry breaking in a hyperbolic space, may hopefully find
further development in more realistic situations with phase
transitions in quark matter under the influence of strong
gravitational fields.

\acknowledgments

We are grateful to A.E. Dorokhov, M.K. Volkov and K.G. Klimenko for
useful discussions. We also appreciate the remarks and helpful
suggestions made by the referee in his report. Two of us (A.V.T. and
V.Ch.Zh.) thank M. Mueller-Preussker for hospitality during their
stay at the Institute of Physics of Humboldt-University, where part
of this work has been done, and also DAAD for financial support.
D.E. thanks the colleagues of the Bogolyubov Laboratory for
Theoretical Physics of JINR Dubna for kind hospitality and the
Bundesministerium f\"ur Bildung und Forschung for financial support.
This work has also been supported in part by the Deutsche
Forschungsgemeinschaft under grant 436 RUS 113/477.

\end{document}